\documentclass[prl,showpacs]{revtex4}
\usepackage{graphicx}
\usepackage{amssymb,amsmath}
\usepackage{textcomp}

\begin{document}

\title{Electronegativity in quantum electronic transport}

\author{R.J. Bartlett$^1$, G. Fagas$^2$, and J.C. Greer$^2$}
\email{jim.greer@tyndall.ie}
\affiliation{$^1$Quantum Theory Project, 
Departments of Chemistry and of Physics, University of Florida Gainesville
FL 32611 USA \\
$^2$Tyndall National Institute, Lee Maltings, Prospect Row, Cork, Ireland}
\date{\today}

\begin{abstract}
Electronegativity is shown to control charge transfer, energy level
alignments, and electron currents in single molecule tunnel junctions, all of 
which are governed by correlations contained within the density matrix. 
This is demonstrated by the fact that currents calculated from the one-electron 
reduced density matrix to second order in electron correlation are identical 
to the currents obtained from the Green's function corrected to second order 
in electron self-energy.
\end{abstract}

\pacs{73.40.Rw,73.63.-b,05.60.Gg,31-25.-v}
\maketitle
\clearpage

Prediction of electron transport across single molecules
requires determination of electronic structure in
the presence of open boundary conditions,
whether using a non-equilibrium statistical or dynamical 
theory~\cite{stat,dyn}. 
Statistical approaches concentrate directly on 
the non-equilbrium density matrix, whereas 
if the time evolution for a system driven from
equilibrium is followed, attention is usually focused on the 
non-equilibrium Green's functions (NEGF) describing electron 
propagation. Treating electronic
structure within transport theories requires an understanding
of the intriguing and challenging to calculate effects of
electron correlations. As exact approaches are limited to
model systems or nanostructures with a small number of
electrons, attention has focused on improving addition
spectra both in the independent electron approximation~\cite{single}
and by many-body treatments through the GW scheme~\cite{many}.
In the following, we 
consider correlation corrections to independent particle models 
and relate conditions on the one-electron Green's function 
and reduced density matrix for
calculation of currents within non-equilibrium 
theories. Correlation corrections to the density matrix
are shown to correspond to improving ionization
potentials (IPs) and electron affinities (EAs) 
given by Green's functions.
This leads to a discussion of electron currents
in terms of electronegativity: the impact of 
the  electronegativity 
on charge transfer, energy level alignments, and current
magnitudes is determined.

Electron currents may be calculated from the one-electron reduced 
density matrix~\cite{Fre90, DeG04} as
\begin{equation}
J({\bf r}) = 
\frac{1}{2i}[\nabla_{\bf r}-\nabla_{\bf r^\prime}]
             \rho({\bf r,r^\prime})|_{\bf r^\prime=r},
\end{equation}
with $J$ the current density, ${\bf r}$ a position vector,
and $\rho$ the one-electron reduced density matrix (RDM); atomic
units are implied unless otherwise given.
As the current density operator is a one-body, non-local operator,
it is clearly necessary to develop accurate approximations to the 
RDM to obtain reasonable results. From another viewpoint, calculation 
of the current can also proceed through computation of the one 
electron retarded and advanced Green's functions $G_{r,a}$ and 
application of a Landauer-type formula~\cite{IV_NEGF}:
\begin{equation}
\label{Land}
I = \frac{1}{\pi}\int d\omega\, [f_L(\omega;\mu_L) -f_R(\omega;\mu_R)]
{\rm Tr} [\,\Gamma_L(\omega)\, G_a(\omega)\, \Gamma_R(\omega)\, \Lambda\, G_r(\omega)\, ],
\end{equation}  
with electron energy $\omega$, $\Gamma_{L,R}$ spectral densities, $f_{L,R}$ 
energy distributions with $\mu_{L,R}$ 
chemical potentials in the left (L) and right (R) electron reservoirs,
and $\Lambda$ is the correction due to correlations
weighted by the spectral density of the electrodes and electron-electron 
spectral density on the molecule.
The causal Green's function is related to the RDM via the relation
\begin{equation}
\label{Coulson}
\rho({\bf r,r^\prime}) = \frac{1}{2\pi i} \oint d\omega\, G({\bf r,r^\prime};\omega),
\end{equation}
with the complex integration performed along the Coulson contour. 
We begin by pointing out that the  reduced density matrix obtained from
a many-electron wavefunction corrected to second order in electron correlation is
equivalent to  the reduced density matrix arising from correcting IPs and EAs in 
the Green's function to second order in the electron self-energy~\cite{PiG73}. 

To proceed, the energy operator for a molecule within a tunnel junction is written 
in the form
\begin{eqnarray}
\label{PertLambda}
\hat H(\lambda)= 
\int d{\bf r}\,\hat\psi^\dagger({\bf r})h({\bf r})\hat\psi({\bf r})
+\int d{\bf r}\,d{\bf r^\prime}\hat\psi^\dagger({\bf r}) v_{\rm HF}({\bf r},{\bf r^\prime})\hat\psi({\bf r^\prime})\nonumber\\
+\lambda\,\big[\frac{1}{2}\int d{\bf r}\,d{\bf r^\prime}\hat\psi^\dagger({\bf r})\hat\psi^\dagger({\bf r^\prime}) 
                      v({\bf r},{\bf r^\prime})\hat\psi({\bf r^\prime})\hat\psi({\bf r})- 
\int d{\bf r}\,d{\bf r^\prime}\hat\psi^\dagger({\bf r}) v_{\rm HF}({\bf r},{\bf r^\prime})\hat\psi({\bf r^\prime})\big],
\end{eqnarray}
with $v$ the electron-electron interaction on the molecular region, $v_{\rm HF}$ the Hartree-Fock potential and 
$\hat\psi^\dagger,\hat\psi$ are second quantized electron field operators. It is assumed that
the Fock equations have been solved with electrode self-energies
$\Sigma^{L,R}$ to describe the interaction between the molecular region electrons and
electrons in the reservoirs; external potentials are also included in the Fock operator. 
For $\lambda =0$, the  Hamiltonian reduces to the Fock operator
\begin{equation}
\hat H(0) = \hat F = \sum_p \epsilon_p \, \hat a^\dagger_p \hat a_p
\end{equation}
with $\hat a^\dagger, \hat a$ creation and annihilation operators for Hartree-Fock states.
For $\lambda = 1$, the many-electron Hamiltonian is restored.
A perturbation expansion in $\lambda$ is written for the many-electron wavefunction: 
\begin{equation}
\label{pert_Psi} 
|\Psi> = |\Psi^{(0)}> + \lambda |\Psi^{(1)}> +\lambda^2 |\Psi^{(2)}> + \ldots\,\,\, .
\end{equation}
For our choice of $0^{th}$ order approximation, Brillouin's theorem insures that 
the first order wavefunction consists of only double electron 
excitations, on the other hand the second order term includes single through quadruple 
excitations. From
\begin{equation}
\rho({\bf r,r^\prime}) = <\Psi |\hat\psi^\dagger({\bf r^\prime})\hat\psi({\bf r}) |\Psi>,
\end{equation}
to first order in $\lambda$ the correction to the 0$^{th}$ order density 
matrix vanishes~\cite{Dav72}. The density matrix to second order is 
\begin{equation}
\rho \approx \rho^{(0)} + \lambda^2 \rho^{(2)}.
\end{equation}
The RDM may be represented as an infinite expansion over single electron states $\phi$
\begin{equation}\label{RDMexp}
\rho({\bf r,r^\prime}) = \sum_{pq} \rho_{pq} \phi_q^*({\bf r^\prime}) \phi_p({\bf r}). 
\end{equation}  
Explicit calculation of the density matrix coefficients from eq.~\ref{pert_Psi} through 
second order in $\lambda$ yields
\begin{eqnarray}
\label{rho_2nd_order}
\rho_{i\, j} &=& \delta_{i\,j}
-\frac{1}{2}\sum_{abk} \frac{ <ab||ik><jk||ab>}
{(\epsilon_i+\epsilon_k-\epsilon_a-\epsilon_b)
 (\epsilon_j+\epsilon_k-\epsilon_a-\epsilon_b)}   \\
\rho_{a\, b} &=& \frac{1}{2}\sum_{ijc} 
\frac{ <ij||ac><bc||ij>}
{(\epsilon_i+\epsilon_j-\epsilon_c-\epsilon_a)
 (\epsilon_i+\epsilon_j-\epsilon_c-\epsilon_b)}   \\
\rho_{i\, a} &=& \frac{1}{2}\sum_{abj} 
\frac{ <ab||ij><aj||ab>}
{(\epsilon_i-\epsilon_a)(\epsilon_i+\epsilon_j-\epsilon_a-\epsilon_b)}
-\frac{1}{2}\sum_{ijb} 
\frac{ <ij||ib><ab||ij>}
{(\epsilon_i-\epsilon_a)(\epsilon_i+\epsilon_j-\epsilon_a-\epsilon_b)} 
\end{eqnarray}
with $<pq||rs> = <pq|v|rs> - <pq|v|sr>$. We use the convention whereby indices 
$i,j,k,\ldots$ label occupied, $a, b, c, \ldots$ label unoccupied, and 
$p,q, r, \ldots$ are used to label general (occupied or unoccupied) states 
in $|\Psi^{(0)}>$. 

Transmission resonances are given through the poles of the 
Green's functions and can be identified as IPs and EAs. Hence, 
it is reasonable to assume that if an independent particle picture is 
chosen to optimize IPs and EAs,
it follows that prediction of currents from the NEGF
approach will be improved. In this context, a model
for transport is measured in terms of reproducing 
the molecular electronegativity.
It is known that introduction of correlation
corrections beyond independent particle models for the Green's function improves 
the predicition of IPs and EAs. The Green's
function with second order self-energies has been studied by Pickup and 
Goscinski~\cite{PiG73} 
leading to the following approximation 
\begin{eqnarray}
\label{correctedG}
[G^{(2)}(\omega)]^{-1}_{pq} &=& [G^{(0)}(\omega)]^{-1}_{pq} + \Sigma^{(2)}(\omega)_{pq} \nonumber\\
                           &=& (\omega -\epsilon_p)\delta_{pq}
-\frac{1}{2} \sum_{iab} \frac{<ab||pi><qi||ab>}{\omega +\epsilon_i-\epsilon_a-\epsilon_b} 
-\frac{1}{2} \sum_{ija} \frac{<ij||pa><qa||ij>}{\omega +\epsilon_a-\epsilon_i-\epsilon_j}.
\end{eqnarray}
The lowest order improvement to Koopmans' IPs and EAs are 
obtained from the poles of the diagonal elements of $G(\omega)$. 
It is found the self-energy
corrects Koopmans' IP
$\epsilon_i$ through terms describing orbital relaxation
and pair correlations;
a similar interpretation holds for corrections 
to the EAs~\cite{PiG73}. Within this
approximation, it is also possible to determine the density matrix directly
from eq.~\ref{Coulson}; the resulting density matrix coincides {\it exactly}
with the density matrix calculated from eq.~\ref{pert_Psi} through $O(\lambda^2)$. Hence 
calculating the density matrix through second order in electron correlation
and correcting IPs and EAs with second order self-energies $\Sigma^{(2)}$ 
will lead to the same predictions for electron current. For moderate
electron correlations, improving spectra for independent particle models or 
explicitly including correlations in the RDM are equivalent. 

Recently a criterion for selecting an independent particle model for quantum 
electronic transport was given as the set of single particle states yielding
an approximate density matrix with maximal overlap to the exact RDM~\cite{FDG06}.
The single electron states diagonalizing the RDM are natural 
orbitals (NOs)~\cite{Low55} and their eigenvalues $\rho_i$ are known as
natural occupations.  If one asks what is the best finite expansion approximation
${\tilde \rho}$ to the exact RDM
\begin {equation}
\label{leastsq}
\Delta \rho = \int | \rho - {\tilde \rho} |^2 d{\bf r} \, d{\bf r^\prime},
\end{equation}
it is found that including the first $n$ natural orbitals with the largest occupancies for 
a truncated expansion eq.~\ref{RDMexp} fulfills the least squares condition~\cite{Dav72}. 
We consider the couplings between density matrix coefficients in eq.~\ref{rho_2nd_order} 
by writing
\begin{equation}
\label{NOmatrix}
\rho = \left[
\begin{array}{cc}
\rho_{i\, j} & \rho_{ i\, a} \\
\rho_{a\, i} & \rho_{ a\, b}
\end{array}
\right],
\end{equation}
with $(i\, j)$, $(a\, b)$, and $(i\, a)$ denoting occupied-occupied, unoccupied-unoccupied,
and occupied-unoccupied spaces respectively, with occupations referred to the 
$0^{th}$ order wavefunction. The natural orbitals to second order in electron
correlation are given by the eigenfunctions of eq.~\ref{NOmatrix}.
Constructing the ``best" independent particle picture in the sense of eq.~\ref{leastsq} 
implies occupying a single Slater determinant by the first $n_e$ natural orbitals.
We have previously shown numerically 
that a single determinant composed of the largest occupation number NOs can lead to essentially 
the same results as a full many-body treatment for tunneling through alkanes~\cite{FDG06}.
For a single determinant approximation, the density matrix is idempotent $\rho^2=\rho$
which occurs since the first $n_e$ occupations are equal to 1 with all others 0.
Hence a measure for the quality of a single determinant approximation is 
how well the eigenvalues of eq.~\ref{NOmatrix} approximate the idempotency 
condition. As the $\rho_{i\, a}$ couplings between the occupied and unoccupied
spaces becomes stronger, the occupations of the $0^{th}$ order states can become significantly
less than unity. From many-body theory it is well understood what this condition implies: a single
determinant or independent particle picture is no longer useful as a 0$^{th}$ order wavefunction.
For weak to moderate correlations, the Green's function approach can achieve improved IPs and EAs by
a low order approximation to the electron self-energy. 
As natural occupancies in the $0^{th}$ order wavefunction become very much less than unity, a 
perturbation expansion about an independent  particle picture loses meaning and 
even higher order corrections
to $|\Psi^{(0)}>$ will not correct IPs and EAs on the molecular region.
In a similar context, this is seen as the failing of the $GW$ 
approximation for systems with multi-determinantal ground states~\cite{PaH07} or
in strongly correlated electron transport~\cite{many}. For strong electron correlations
coupled-cluster theory offers a convenient nonperturbative framework from which higher order 
approximations to the density matrix follow~\cite{BGH05}, alternatively correlated one particle 
methods~\cite{BeB} to infinite order can be chosen to yield correct IPs and EAs.

\begin{figure}
\includegraphics[width=4.2in]{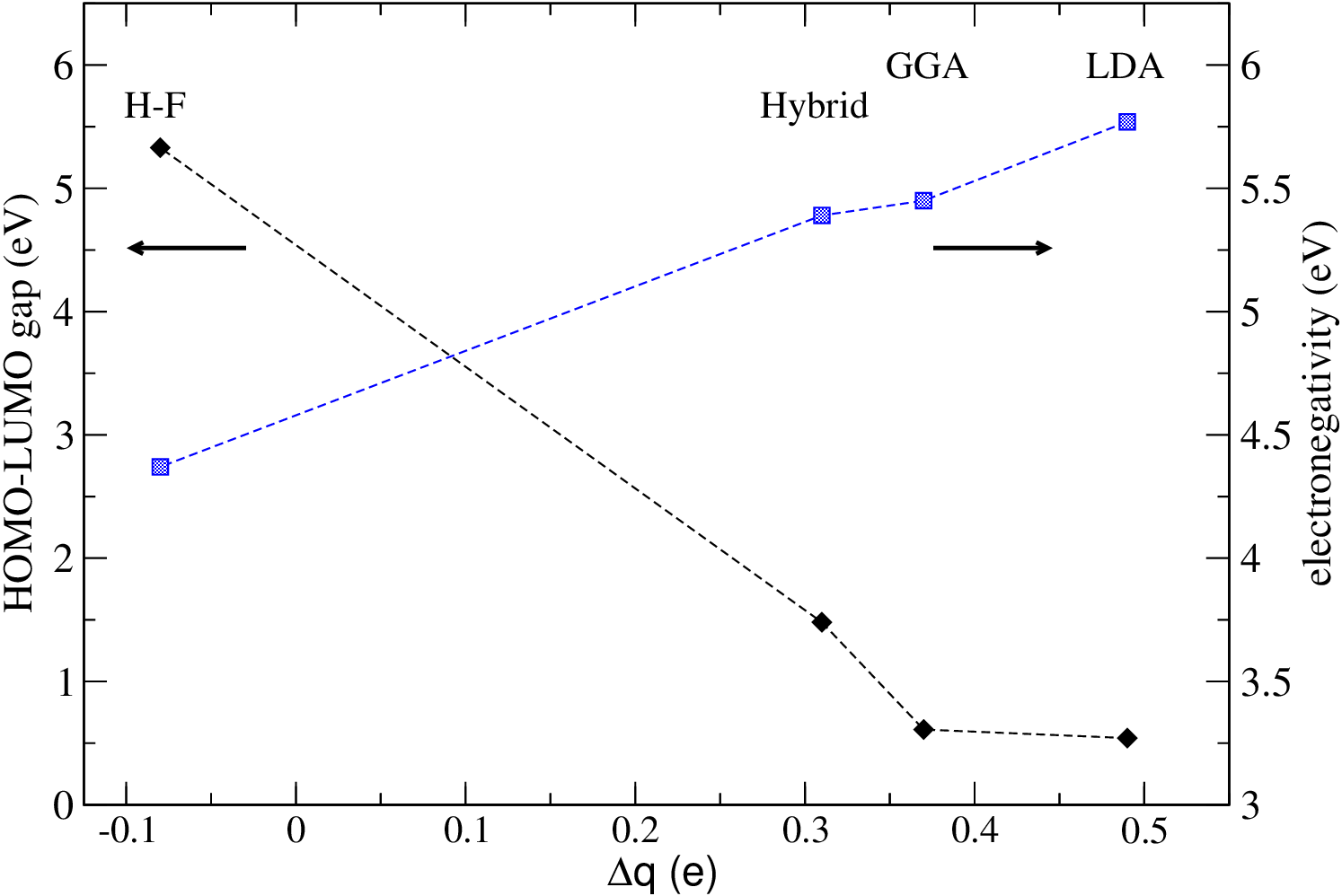}
\caption{Highest occupied and lowest unoccupied (HOMO-LUMO) energy gap
and electronegativity versus charge transfer for hexenedithiol
bonded to linear gold chains. Calculations have been performed with the
TURBOMOLE program system~\cite{TM}. All calculations have been performed using the
auc-cc-pVDZ basis set for carbon \cite{KDH92} and split valence polarized valence
basis for all other atoms, including a sixty electron effective core potential for
the gold atoms~\cite{TM}. Calculations have been performed using the Hartree-Fock 
and density functional theory calculations using hybrid (B3-LYP), generalized
gradient approximation (GGA/PBE), and local density approximation (LDA/PW)
exchange-correlation functionals.
}
\label{fig1}
\end{figure}

The Mulliken electronegativity given as $(IP+EA)/2$ is a useful measure of charge transfer, 
and it is charge transfer 
that determines molecular level alignments relative to electron reservoir energies~\cite{StJ06}. 
Predicting level alignments correctly for molecules bonded between electrodes is essential
for accurate current-voltage characteristics~\cite{XDR01}. In  the Hartree-Fock 
approximation, charge transfer is under-estimated as hybridization to virtual states is weak.
In the local density (LDA) and generalized gradient (GGA) approximations to density 
functional theory (DFT), charge transfer is over-estimated~\cite{RSV96}. These effects are demonstrated
for the case of hexenedithiol bonded between two gold clusters in fig.~\ref{fig1} where 
the highest occupied-lowest unoccupied energy gap in the molecular orbitals (HOMO-LUMO gap) and
molecular electronegativity is
given against charge transfer relative to molecular hexenedithiol. 
For a large HOMO-LUMO gap or weak electronegativity, charge transfer is small. For small HOMO-LUMO 
gaps typical of GGA and LDA, over-estimation of charge transfer is confirmed. Hybrid functionals 
correct the  charge transfer to some extent, but this correction is not systematic~\cite{RSV96}.

\begin{figure}
\includegraphics[width=4.2in]{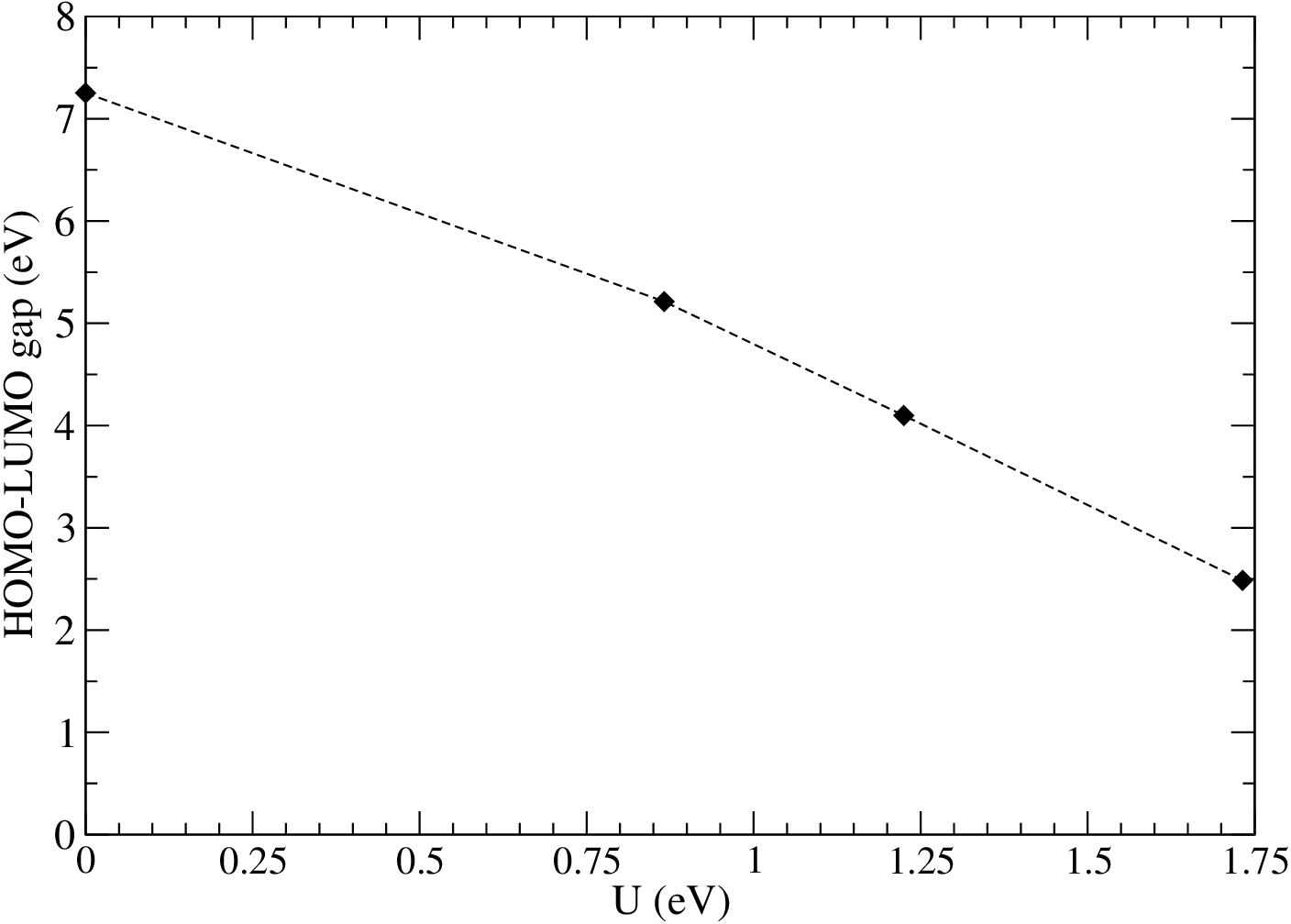}
\caption{HOMO-LUMO gap for the model system defined by eq.~\ref{ModelH} as 
a function of the electron-electron self energy as varied through the 
interaction parameter $U$. The reduction in the gap demonstrates the effect
of electron-electron self-energy on the molecular electronegativity.
}
\label{fig2}
\end{figure}

\begin{figure}
\includegraphics[width=4.2in]{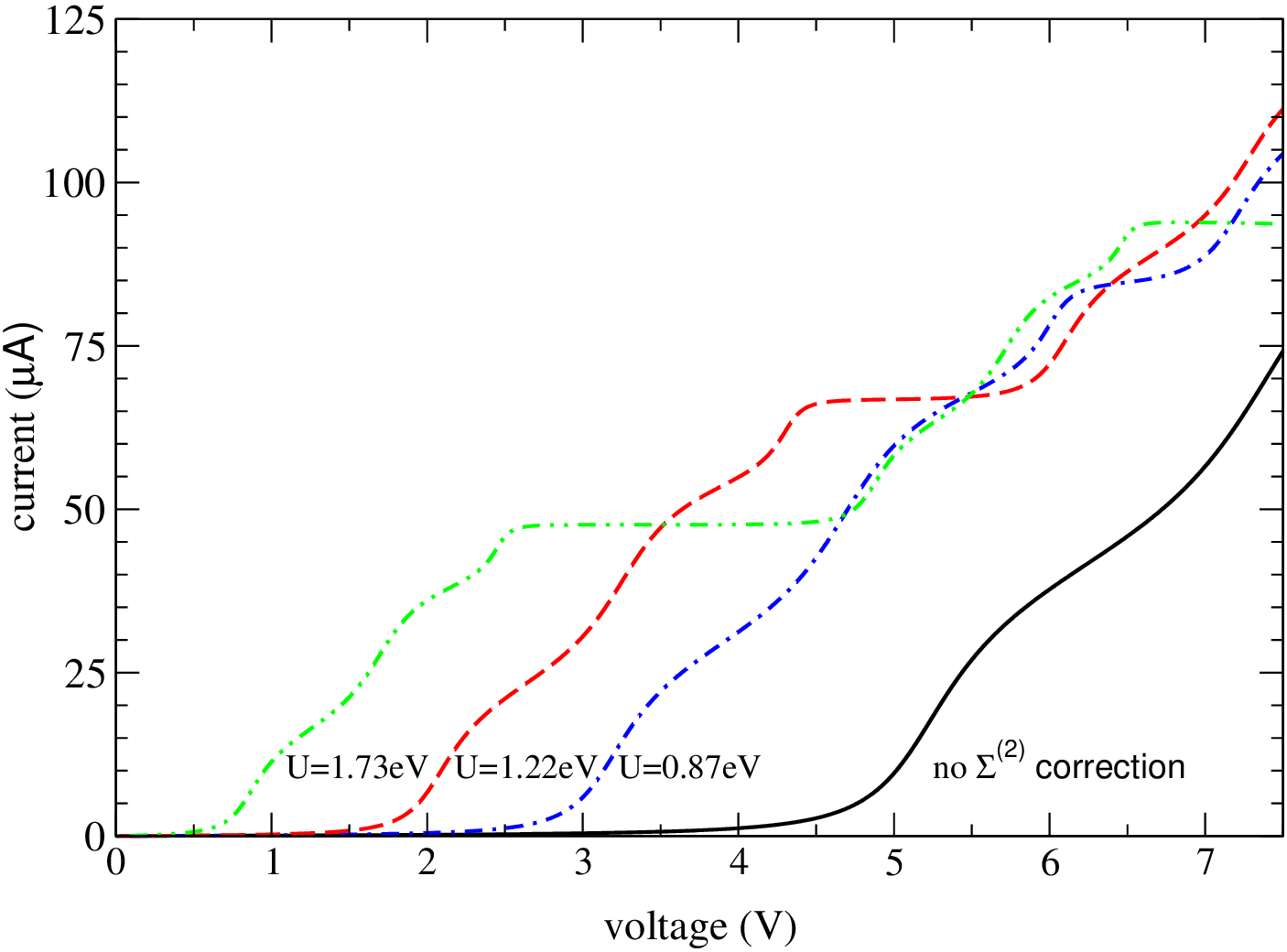}
\caption{Current voltage characteristics for the model Hamiltonian of eq.~\ref{ModelH}.
$\epsilon_M=1.0 eV$, $\epsilon_L=\epsilon_R= \epsilon_{\rm Fermi}=0.0$, $\gamma_M=4.54 eV$,
$\Gamma_M=1.5 eV$, $\gamma_L=\gamma_R=10.0 eV$, 
$\gamma_{LM}=\gamma_{MR}= 2.4 eV$. Electronegativity is modified by varying $U$, with values
as labeled within the figure.
}
\label{fig3}
\end{figure}

We introduce a simple 
correlated model for a molecular chain and investigate the effect of over- and under-estimation
of electronegativity on electron transport. 
We use the following model Hamilitonian for an infinite chain:
\begin{eqnarray}
\label{ModelH}
\hat H = 
&-& \gamma_L  \sum_{n < -3} (\hat c_n^\dagger\,\hat c_{n-1}^{ } + h.c.) 
+             \sum_{n < -3} (\epsilon_L + V_L)\, \hat c_n^\dagger \, \hat c_n^{ }   
 - \gamma_{LM} ( \hat c_{-4}^\dagger \hat b_{-3}^{ } + h.c.)  \nonumber\\
&+&\sum_{n=-3}^{+3} (\epsilon_M+V_n) \, \hat b_n^\dagger \, \hat b_n^{ } 
 - \gamma_M    (\hat b_{-3}^\dagger\,\hat b_{-2}^{ } + \hat b_{-1}^\dagger\, \hat b_{1}^{ } + \hat b_2^\dagger\, \hat b_{3}^{ } 
             + h.c.)
 - \Gamma_M (\hat b_{-2}^\dagger\, \hat b_{-1}^{ } + \hat b_1^\dagger\, \hat b_{2} 
            + h.c.) \nonumber \\
&-& \gamma_{MR} (\hat  b_{+3}^\dagger \hat c_{+4}^{ } + h.c.)
+             \sum_{n > +3} (\epsilon_R + V_R)\, \hat c_n^\dagger \, \hat c_n^{ }       
- \gamma_R    \sum_{n > +3} (\hat c_n^\dagger\,\hat c_{n+1}^{ } + h.c.)
\end{eqnarray}
Six central sites of the chain are labelled -3, -2, -1, 1, 2 ,3 (i.e. there is no $0$ site) and are
treated as the molecular region with $\hat b^\dagger, \hat b$ 
creation and annihilation operators for electrons on the molecule. The electron reservoirs
are described by the atomic sites extending towards the left and right away from the central
molecular sites with creation and annihilation operators $\hat c^\dagger, \hat c$ for the 
reservoir electrons. The site energies are given by $\epsilon_L=\epsilon_R$ and $\epsilon_M$ for
the reservoir and molecular regions, respectively. The volage applied across the molecular junction is
described by the voltages $V_L \ne V_R$ in the reservoirs 
and the voltage drop $V_n$ across the molecular sites is
scaled linearly between the values $V_L$ and $V_R$. The nearest neighbor interactions are 
$\gamma_L=\gamma_R$ within the electrode regions, and there are two molecular site-site interaction
$\Gamma_M$ and $\gamma_M$ representing single and double bonds, respectively, on the molecular region
as a simple model for hexenedithiol, and 
$\gamma_{LM}=\gamma_{MR}$ determine the molecule-electrode couplings. 
The eigenstates of the molecular Hamiltonian are found with the electron-electron
self-energies and exact electrode self-energies are introduced describing
coupling to the electrodes~\cite{HFH06}.
The resulting single electron states are taken as the expansion functions for the correlated version of the model 
obtained from $\hat H_0  \rightarrow \hat H_0 + \hat v$, with $\hat v$ the pairwise perturbation 
interactions about the mean field solution as in eq.~\ref{PertLambda} with $\lambda=1$. 
Current-voltage 
characteristics are calculated using eq.~\ref{Land}. We use a simplified form of the self-energy
such that the interaction matrix elements in eq.~\ref{correctedG} are  approximated as $<pq||rs> \approx U$.
In fig.~\ref{fig2}, the HOMO-LUMO gap for the molecular region is given as
a function of $U$ demonstrating that the electronegativity on the molecular region may be
systematically controlled through the electron-electron self-energies.
The results for the current voltage characteristics from the model are presented in fig.~\ref{fig3}. The 
independent particle or uncorrelated model 
occurs for $U = 0$ and increasing $U$ corresponds to increasing
electron correlations on the molecular region. At $U=0$, currents
at low voltages are much lower than when the $\Sigma^{(2)}$ term is allowed to correct
IPs and EAs; in this case, the highest lying occupied states are too low (IPs too
high) and the lowest lying unoccupied single electron states are too high (EAs too low) with respect
to the Fermi level. Under these conditions
neither occupied or unoccupied states enter into the voltage bias window at low voltages, 
and this level of electronic structure treatment corresponds to a
Hartree-Fock approximation. 
Increasing correlations on the molecular region, the highest occupied states near 
the Fermi level enter the bias window at lower values of voltage, followed by the introduction
of the unoccupied states at higher voltage bias (this sequence is due to the relative
position of the Fermi level relative to occupied and unoccupied states for this example). The 
correlations on the molecular region serve to shift up occupied levels relative to the 
Fermi level leading to reduced IPs, 
whereas increasing correlations systematically lower the lowest lying unoccupied states leading to 
increased EAs.  Increasing correlations continue to reduce the IPs and increase
EAs until eventually electronegativity is under-estimated. The impact on the current-voltage
characteristics is that the molecular levels enter
the bias window at very low values of applied voltage resulting in large current magnitudes. 
Larger values of $U$ correspond to the use of LDA or GGA exchange-correlation potentials 
within DFT where the strong over-estimation of charge transfer is known to occur~\cite{RSV96}. 
Hartree-Fock and Kohn-Sham (using LDA or GGA) orbitals are not appropriate independent particle models for
electron transport due to strong under- and over-estimation of charge transfer, respectively. The
results of fig.~\ref{fig3} clearly show the impact on current voltage characteristics for these two extremes.

Correcting electronegativity is equivalent
to maximizing overlap to the reduced density matrix: this is true to low orders in electron correlation
and of course the correct electronegativity and density matrix are found at the exact many-body solution.
In general, improving descriptions for the RDM and electronegativity with the methods described will lead 
to improved prediction of electron currents in systems with moderate electron correlations. 
The best independent particle picture within this context is a single determinant comprised of natural 
orbitals; any attempt to refine single electron 
models for transport should lead to electron wavefunctions that approximate natural orbitals.  
In the case of Green's function approaches, moderate electron correlations
imply the need to include electron-electron self-energies to describe quasi-particle propagation.
For strong correlations, a single determinant wave function is not an 
adequate approximation to predict IPs and EAs and perturbation corrections about a 
single reference state fail- thus complicating treatment of molecular junctions with Green's function approaches. 
However, in all cases, from weak to strong correlations, the criterion to maximize overlap to the exact reduced 
density matrix leads to improved predictions for electron currents.   

{\bf Acknowledgments} This work was supported by a Science Foundation Ireland.

\clearpage

\clearpage

\clearpage

\end{document}